# Criteria to observe single-shot all-optical switching in Gd-based ferrimagnetic alloys

Wei Zhang[1,2,3], Julius Hohlfeld[3], Tian Xun Huang[3], Jun Xiao Lin[3], Michel Hehn[3], Yann Le Guen[3], Jude Compton-Stewart[3], Gregory Malinowski[3], Wei Sheng Zhao[1,2*], Stéphane Mangin[3*]

[1]National Key Lab of Spintronics, Institute of International Innovation, Beihang University, Yuhang District, Hangzhou, 311115, China

[2]Fert Beijing Institute, BDBC, Beihang University, Beijing 100191, China

[3]Université de Lorraine, CNRS, IJL, F-54000 Nancy, France

**Abstract**

Single-shot all-optical helicity-independent switching (AO-HIS) induced by a femto-second laser pulse has been mainly reported in Gadolinium based rare earth-transition metal (RE-TM) alloys such as GdFeCo or GdCo, but the mechanism leading to magnetization switching is a hotly debated topic. Here, we elaborate on a large number of $Gd_yRE_{1-x-y}Co_x$ (RE = Dy, Tb, Ho) alloys to tune various magnetic parameters in order to define what the criteria are for observing AO-HIS in such systems. The state diagrams show that two laser fluences thresholds must be considered: the fluence which induces the single laser pulse switching ($F_{Switch}$) and the fluence at which the material breaks into a multi-domain state ($F_{Multi}$). Those two fluences are shown to behave very differently as a function of the material properties and the laser pulse duration. Taking into account the parameters defining the conditions for which multi-domain states are created and considering only the angular momentum transfer from the Gd sublattice to the rest of the system explains in large our experimental results. The importance of the compensation in the ferrimagnetic alloys is also discussed. We believe the defined criteria will be an important tool for designing new ultra-fast spintronic devices based on all optical switching.

# I. INTRODUCTION

As a powerful way to sense, store and process digital information, magnetics-based technologies [1-3] undoubtedly promote the development of modern society. Controlling and manipulating magnetic configurations with simultaneously higher energy-efficiencies and at shorter timescales is not only a new challenge, but also an urgent issue for evolving information technology [4, 5]. As the fastest external stimulus, the use of light is a promising approach to enhance the speed of data processing. As a prominent technique of light-induced magnetic manipulation, single-shot ultra-fast all-optical helicity-independent switching (AO-HIS) [6] is of crucial importance for generating smaller, faster and more energy efficient information technology. AO-HIS taking place on a picosecond timescale has been demonstrated experimentally in a rather limited number of materials. First, in Gadolinium-transition metal (Gd-TM) alloys (GdFeCo or GdCo) using a femtosecond laser pulse without any applied magnetic field [7-11]. In those heavy rare earth – transition metal (RE-TM) ferrimagnetic alloys, the magnetization of the TM sublattice is antiferromagnetically exchange coupled to the magnetization of the RE. Consequently, at certain compositions, a compensation temperature exists at which the net magnetization of the alloy is zero. Recently, breakthrough experiments have shown that single laser pulse switching can be achieved in multilayer structures including Gd/Co [12], Tb/Co [13] synthetic ferrimagnets, or the rare earth-free Heusler alloy $Mn_2Ru_xGa$ [14]. Note that single pulse switching in Tb/Co presents a very different switching mechanism, which is not ultra-fast and shows a very different state diagram [15].

Single pulse, ultra-fast, AO-HIS has been observed only in alloys and multilayers where two magnetization sublattices are antiferromagnetically coupled. Furthermore, with the known exception of $Mn_2Ru_xGa$ [14], the presence of Gadolinium and an alloy concentration close to compensation is required. Indeed, in case of GdFeCo or GdCo alloys, several experimental and theoretical works have demonstrated that AO-HIS was possible only for a limited composition range around the compensation point [6, 9, 16-18]. Note that AO-HIS for TbCo alloys was predicted but has never been demonstrated [19]. This limits the range of materials which can be used for possible all-optical data storage applications. Moreover, it is still not clear (i) if we can reduce or remove Gd in the RE-TM alloys and still observe AO-HIS and (ii) if being close to compensation is the main material criteria for observing AO-HIS as suggested by the theories [6, 18] or if other criteria must be met?

Fortunately, exploring the parameter spaces, such as material properties and laser pulse characteristics (fluences F and pulse durations τ) for the observation of AO-HIS, is helpful to determine the criteria to be fulfilled to observe ultra-fast, single-pulse, AO-HIS. The effect of the above parameters on AO-HIS are summarized by building state diagrams showing the final magnetic state of the sample after a single laser pulse as a function of F and τ. As described by Wei *et al*. [16], the state diagram for AO-HIS is shown to have a triangular shape and three quantities corresponding to the three edges of this triangle can be determined and discussed. $F_{Switch}^{Min}$ is the minimum fluence at which AO-HIS can be observed (usually for the shortest pulse duration), $F_{Multi}$ is the fluence defining the transition from AO-HIS to a multi-domain state (almost independent on the pulse duration), and $\tau_{max}$ is the maximum pulse duration for which AO-HIS can be observed. Gorchon *et al*. [20] demonstrated that $\tau_{max}$ can be as long as 15 ps in GdFeCo. This challenged the accepted driving mechanism, being the ultrafast angular momentum transfer between RE and TM sublattices, since the initial experiments for AO-HIS used ultrashort (~100 fs-long) optical pulses. Later, Davies *et al*. [9], by modelling the ferrimagnet as two coupled macrospins, and Jakobs *et al*. [6] using an atomistic spin dynamics model, were able to reproduce that a large $\tau_{max}$ works for AO-HIS only if the alloy composition is close to compensation. However, all those studies have only been carried out for GdFeCo alloys, leaving the development of the state diagrams of other GdRECo alloys, such as RE = Dy, Tb, Ho largely undocumented. This has also significantly limited our ability to establish a full criterion for AO-HIS observation.

In this work, we have grown series of 10nm thick $Gd_yRE_{1-x-y}Co_x$ thin film alloys (RE = Dy, Tb, Ho) in order to determine the influence of composition on AO-HIS. We could first demonstrate AO-HIS for a much larger range of Co and RE concentrations than those seen in pure GdCo or GdFeCo alloys. The possibility to introduce rare earth such as Dy, Tb and Ho in the alloys allows to increase the magneto-crystalline anisotropy and consequently increase the thermal stability of potential devices. Moreover, the amount of Gd needed to observe AO-HIS could be as low as 1.5 %. We demonstrate that the characteristic $F_{Switch}$ versus τ and $F_{Multi}$ versus τ extracted from triangular state diagrams for AO-HIS vary differently with the alloy's magnetic parameters such as $M_s$ (saturated Magnetization), $J_{eff}$ (effective exchange coupling), $K_u$ (Magneto-crystalline Anisotropy Constant) and the amount of Gd. We could then define the criteria needed to be fulfilled to observe

AO-HIS. We could demonstrate why in many cases being close to compensation is essential to observe magnetization switching under a single femto-second laser.

## II. EXPERIMENTAL RESULTS

Fig. 1(a) shows the whole series of $Gd_yDy_{1-x-y}Co_x$ alloys we studied in this work, with x varying between 61% and 80%. In this work, we mainly focus on the samples with perpendicular magnetic anisotropy (PMA) as our experimental optical setup allows the measurement of the out of plane magnetization component. Among the samples with PMA, two kinds of magnetic states can be distinguished after excitation by a single femtosecond laser pulse, *i.e.,* the multi-domain state represented by the dark red region, and the deterministically switched state in blue marked within the white line. Note that for a few Gd rich samples located in the red region, the magnetization is in plane, the orientation of which our experimental set-up does not allow us to observe. For all the $Gd_yTb_{1-x-y}Co_x$ and $Gd_yHo_{1-x-y}Co_x$ alloys samples, a very similar mapping is observed as shown in Fig. S1(a) and (b) in Note.1 of the Supplementary Materials. For $Gd_yDy_{1-x-y}Co_x$ alloys, the concentration of Co (x) and Gd (y) allowing AO-HIS covers from x=63% to x=78%, and y=1.5% to y=26%. The engineering of RE-TM alloys with a large enough compositional window showing AO-HIS is highly desirable since the magnetic properties ($M_s$, $K_u$, $T_c$) can then be strongly selected for [21, 22]. In Fig. 1(b), we summarized the range of Co compositions, and the minimum Gd compositions needed to show AO-HIS. By introducing other rare earth elements Tb/Dy/Ho, the windows of Co concentrations (Δx) showing AO-HIS is Δx =17% / 15% /15%, respectively, which is much larger than that for pure GdCo alloys (2%), even larger than that in GdFeCo alloys (5%) reported by Wei [16] and Davies *et al*. [9]. Moreover, we found that a minimum Gd composition around 1.5% is enough to sustain the AO-HIS in $Gd_yRE_{1-x-y}Co_x$ alloys, a value lower than the results in GdTbCo alloys reported previously by Ceballos *et al.* [23].

### 1. Samples' Magnetic Properties

The magnetization of the samples, namely glass/Ta(5 nm)/Pt(3 nm)/$Gd_yRE_{1-x-y}Co_x$ (10 nm)/Pt(5 nm) have been measured by applying a field perpendicular to the film plane to determine if the sample magnetization is in plane or out of plane (Fig. 1(a)). Fig. 2(a) shows the hysteresis loops obtained using the magneto-optic Kerr effect (MOKE) at room temperature with a magnetic field applied perpendicular to the film plane for various $Gd_yDy_{1-x-y}Co_x$ alloys when x = 72%. These square loops with 100% of remanence indicate a well-defined strong perpendicular magnetization anisotropy (PMA) of the samples. As compared with that for x = 72% in Fig. 2(a), the polarity of the hysteresis loops in Fig. 2(b) shows a reversed sign at concentration value of x = 74%. Given that the MOKE technique senses predominantly the Co subnetwork, the change of the signal's polarity suggests the magnetization compensation point $x_{comp}$ is between x = 72% and 74% at room temperature, whatever the Gd concentration. This is consistent with the results in Fig. 1(a), where the boundary between Co-dominant and RE-dominant samples is around x = 72%. Usually, the net magnetization approaches zero and the coercivity diverges at the magnetization compensation point [24]. In Fig. 2(c) and d, a color scale from red to blue allows us to map the evolution of the coercivity ($H_c$) and saturation magnetization ($M_s$) as a function of the Co and Gd concentration in $Gd_yDy_{1-x-y}Co_x$. This demonstrates that the magnetic properties can be strongly affected by compositional changes. Obviously, the maximum coercivity and minimum saturation magnetization of $Gd_yDy_{1-x-y}Co_x$ occurs around x = 72%, simultaneously. The presence of a shaded region highlights the compensation concentrations in these alloys in Fig. 2(c) and (d). Very similar behavior is observed for Tb and Ho based $Gd_yRE_{1-x-y}Co_x$ alloys. Even though the magnetization of those alloys reaches zero between x = 72% and 74%, the coercivity increases by replacing the Gd by one of the other Rare-earth. This is certainly due to the fact that the magneto-crystalline anisotropy $K_u$ is higher when Gd (an S ion) is replaced by another RE (Tb, Dy or Ho). In those $Gd_yRE_{1-x-y}Co_x$ alloys, the Curie Temperature, $T_C$, and consequently the effective exchange coupling $J_{eff}$, increases when the Co concentration increases [21,22]. Furthermore, for a constant Co concentration, $T_C$ will be higher for GdCo alloys than for TbCo alloys, which will in turn have higher $T_C$s than DyCo alloys and so on for HoCo alloys. In conclusions, with the three-sample series of $Gd_yTb_{1-x-y}Co_x$, $Gd_yDy_{1-x-y}Co_x$, and $Gd_yHo_{1-x-y}Co_x$ alloys we will be able to tune parameters such as $M_s$, $T_C$, $J_{eff}$, $K_u$ and study their effect on the observed AO-HIS behaviors.

## 2. Single-shot AO-HIS in $Gd_yRE_{1-x-y}Co_x$ alloys

The experimental set up used to study single-shot AO-HIS is sketched in Fig. 3(a). Pictures are taken before and after each laser shot, to allow for a background image correction. The contrast is given by the magnetization projection along the direction perpendicular to the film plane. To ensure a well-defined initial state, the film is saturated using an external field whose exact value depends on the coercivity of the samples. We estimate that the picture is taken 1 second after the laser pulse.

In Fig. 3(b), (c) and (d), we focus on GdDyCo samples near the compensation point ($x_{comp}$), i.e., $Gd_yDy_{1-x-y}Co_x$ with x = 72%, while y varies from 1.5% to 23%. A toggle magnetization switching similar to the one observed for GdCo alloys is shown in Fig. 3(b) after 5-subsequent laser pulses with laser fluence F = 3.2 mJ/cm$^2$. Interestingly, deterministic reversal is achieved, even when the Gd concentration is as low as 1.5% (as much as 26.5% of Dy). The same result is obtained for Tb and Ho as shown in Fig. S2, where similar toggle magnetization switching is also recorded for both $GdTbCo_{74}$ and $GdHoCo_{72}$ alloys. Here, the maximum concentration of Tb (24.5%), Dy (26.5%), and Ho (26.5%) allowing for AO-HIS in GdRECo alloys are higher than the previously reported value of 18% for $Gd_4Tb_{18}Co_{78}$ [23]. As one can see from Fig. 1(a), the minimum value of y (Gd concentration) has to be increased gradually to observe the switching when x (Co concentration) is moving away from the $x_{comp}$. In addition to the demonstration of the minimum Gd concentration needed to show AO-HIS, the minimum Co concentrations to sustain AO-HIS are 65%, 63%, and 61%, for $Gd_yTb_{1-x-y}Co_x$, $Gd_yDy_{1-x-y}Co_x$ and $Gd_yHo_{1-x-y}Co_x$ alloys, respectively. The corresponding magneto-optical images after 5-subsequent laser pulses are shown in Fig. S3 for $Gd_{26}Tb_9Co_{65}$, $Gd_{23}Dy_{14}Co_{63}$, and $Gd_{26}Ho_{23}Co_{61}$, where one can see a clear toggle magnetization switching.

**3. AO-HIS state diagrams as a function of the Gd, RE (Tb, Dy or Ho) and Co concentrations**

In this section, the state diagrams showing the magnetic states achieved after a single laser pulse as a function of pulse duration and fluence is investigated carefully, in order to optimize the laser parameters and the energy needed to induce magnetic switching. We start with the $Gd_{23}Dy_5Co_{72}$ sample, in which a large amount of Gd guarantees AO-HIS for a relatively large region of fluence. Fig. 3(c) shows the images obtained for various laser fluences with pulse duration ranging from 50 fs to 5 ps. At first glance, the window of laser fluence allowing magnetization switching strongly depends on the pulse duration. At the shortest pulse duration τ = 50 fs, this fluence window is the widest, narrowing as the pulse duration increases.

The state diagram in Fig. 3(d) is built by gathering the values of the threshold Fluence (F) and Pulse Duration (τ) ) at which AO-HIS ($F_{Switch}$ (τ)) and multi-domain state ($F_{Multi}$ (τ)) are observed, indicated by solid black dots and open dots, respectively. Since $F_{Switch}$ (τ) and $F_{Multi}$ (τ) evolve linearly with τ, the state diagram is then defined by two (red) lines. When the laser fluence used is lower than $F_{Switch}(τ)$, the system doesn't respond to the laser pulse and there is no switching. This corresponds to the pink region in Fig. 3(d). When the fluence is larger than $F_{Multi}(τ)$, a multi-domain state is created after the pulse, showing no AO-HIS effect. By a first approximation, $F_{Multi}$ is independent of the pulse duration, while $F_{Switch}(τ)$ increases linearly with τ. We could then use $F_{Switch}(τ) = kτ + F_{Switch}^{Min}$, where $F_{Switch}^{Min}$ is the minimum fluence needed to switch, experimentally obtained for the shortest pulse duration ( τ = 50 fs ). $F_{Switch}(τ)$ and $F_{Multi}(τ)$ then define the grey triangular area showing AO-HIS. $τ_{max}$ is therefore given by $F_{Switch}(τ_{max}) = F_{Multi}(τ_{max})$ and defines the maximum pulse duration above which AO-HIS cannot be observed, whatever the fluence. $τ_{max}$ is then defined by three parameters: the slope $k$, $F_{Switch}^{Min}$ and $F_{Multi}$.

The evolution of the state diagram as a function of the Gd concentration (as well as the concentration of Dy), when the Co concentration is fixed to 72% is shown in Fig. 4(a). All the state diagrams share a similar shape and we can still consider that $F_{Switch}(τ)$ evolves linearly with τ whereas $F_{Multi}$ is constant. For fixed 72% Co concentration, when decreasing the amount of Gd, the maximum pulse duration $τ_{max}$ and the fluence range at which AO-HIS is observed shrinks. Fig. 4(b) quantitatively presents $F_{Switch}^{Min}$, $F_{Multi}$, and $k$ as a function of the Gd (Dy) concentrations derived from the state diagrams in Fig. 4(a). By increasing the amount of Gd, $F_{Switch}^{Min}$ decreases, whereas $F_{Multi}$ and $k$ increases, increasing $τ_{max}$. In Fig. S4(a) and (b), the evolution of the shape of state diagrams as a function of Co concentrations, for constant Gd and Dy are presented. For details, please refer to the descriptions in Note. 4 of Supplementary Materials. Based on these state diagrams, we could extract the evolution of $F_{Switch}^{Min}$, $F_{Multi,}$ $k$ in Fig. 4(c) and (d), when the concentrations of Gd and Dy are fixed, respectively. In Fig. 4(c), one can see that $F_{Switch}^{Min}$ increases as the Co concentrations increases while the slope $k$ decreases. $F_{Multi}$ exhibits a peak close to the compensation concentration.

## 4. Criteria to observe AO-HIS – influence of magnetic and laser parameters

From the above results, the obvious criteria to observe AO-HIS is to have $F_{Switch}(τ) < F_{Multi}(τ)$. It is true when changing the material concentrations (For example, in Fig. 4(c) and (d)) but also when increasing the pulse duration (Fig. 3(c) and (d)). This is nicely evidenced on Fig. 5 where the evolution of $F_{Switch}$, $F_{Multi}$ and Ms as a function of Co concentration is shown for a fixed Gd concentration (~18%) and for a pulse duration τ = 50 fs. The evolution of $F_{Multi}$ is not monotonic and depends strongly on the alloy concentration. It is a maximum at the compensation point where Ms is minimum (~ 50 emu/cm$^3$), indicating an inverse relationship between Ms and $F_{Multi}$. On the other hand, $F_{Switch}$ is varying monotonically with the Co concentration. $F_{Switch}$ doesn't seem to be affected by the compensation concentration suggesting that compensation is not required and is not helping AO-HIS, however, it increases $F_{Multi}$ which allows for the fulfillment of the $F_{Switch} < F_{Multi}$ criteria.

Let's now try to explain the evolution of $F_{Switch}^{Min}(\tau)$ and $F_{Multi}(\tau)$ and consequently determine what are the requirements to observe AO-HIS. Starting with $F_{Multi}(\tau)$, we know that PMA samples tend to break into domains due to the dipolar field [25-27]. From Ref. [26] and [27], we can define a dimensionless parameter $\Delta = \frac{D_0}{t}$, where $t$ is the sample thickness and $D_0 = \frac{\varepsilon}{E_{dem}}$ is the dipolar length. $\varepsilon = 4\sqrt{K \cdot J}$ is the Domain Wall energy per surface unit and $E_{dem} = 2\pi M_s^2$ is the demagnetizing energy per unit volume. As discussed in [26] and [27], if $\Delta = \frac{2\sqrt{K_u \cdot J_{eff}}}{\pi M_s^2 t}$ is small, the system will easily break into small domains and $F_{Multi}$ will be small whereas if $\Delta$ is large, the system will tend to stay saturated in large domains and $F_{Multi}$ will be bigger. Consequently, $F_{Multi}$ depends on Ku, $T_C$, Ms and t.

If we consider $F_{Switch}$, most of the results described above can be understood by considering the switching as the action of the spin current generated by the demagnetization of the Gd sublattice only ($j = \frac{dM_{Gd}}{dt}$) transferring its angular momentum to the rest of the system when reaching the ordering temperature of the sample ($T_C$) in accord with Ref [29]. This hypothesis is based on the results obtained in Gd-based systems [30-32] and the model developed by Q. Remy *et al.* [33] for Gd-based spin valve structures. This agrees with the results obtained by Choi [34] suggesting the spin current generated by the Gd sublattice can interact with the transition metal when the latter is demagnetized. The $F_{Switch}$ threshold fluence will be reached when enough spin current is generated and when the spin temperature of the alloy is close enough to $T_C$. In those conditions, to reduce $F_{Switch}$, one needs some combination of an increase to the Gd concentration ($y_{Gd}$) or a decrease to the pulse duration, $T_C$, or thickness of the layer, t. Consequently, $F_{Switch}$ depends on $y_{Gd}$, $\tau$, $T_C$ and t. There are certainly two other threshold fluence which would have to be considered, the one for which the sample is demagnetized for a long period of time, also resulting in a multi-domain state and the one which will damage the sample or change its magnetic properties.

In the rest of this paper, we will check if the criteria ($F_{Switch}(\tau) < F_{Multi}(\tau)$) and the expected evolution of $F_{Switch}(\tau)$ and $F_{Multi}(\tau)$ can explain the behaviour of the 300 samples involved in the present study. The layers thicknesses of the sample have been kept constant, however, all the other magnetic parameters ($M_s$, $K_u$, $J_{eff}$) mentioned above have been tuned by changing the alloys concentrations and the nature of the rare-earth (Tb, Dy or Ho).

Let's consider the study presented in Fig. 5 obtained from a series of 10 nm thick GdDyCo alloys, with a constant Gd concentration around 18% using a $\tau = 50$fs pulse duration (i.e., t, $y_{Gd}$ are kept constant). From the above discussion $F_{Switch}$ should only depend on $T_C$ and $F_{Multi}$ on $K_u$, $T_c$ and $M_s$. The monotonical increase of $F_{Switch}$ with the Co concentration can then be understood by the fact that $T_C$ increases with the Co concentration. The maximum value of $F_{Multi}$ at compensation can be explained by the minimum value of Ms at this point. We can then argue that the evolution of Ku and $J_{eff}$ have little influence on $F_{Multi}$ as they only weakly vary compared to Ms and because, in the above expression of $\Delta$, Ms is present at a power of 2 whereas $K_u$ and $J_{eff}$ are at a power of ½. From Fig. 5, we can clearly conclude that if AO-HIS cannot be observed for concentration larger than 78% and smaller than 63 %, it is because for those cases $F_{Switch}(\tau) > F_{Multi}(\tau)$.

Considering that the two threshold fluences can be written as follows - $F_{Switch}(\tau) = k\tau + F_{Switch}^{Min}$ and $F_{Multi}(\tau) = $ Constant – Fig. 6 shows the evolution of the four parameters $F_{Multi}$, $F_{Switch}^{Min} = F_{Switch}(50fs)$, $k$ and $\tau_{max}$ as a function of the Co and Gd concentrations. From Fig. 6(a), Fig. 4(c) and Fig. 4(d) one can clearly see that $F_{Multi}$ is dominated by the value of the alloy's net magnetization which leads to a maximum value at compensation. We mentioned that $F_{Multi}(\tau)$ is almost independent of the pulse duration $\tau$ but some state diagrams show a small decrease of $F_{Multi}$ when increasing the pulse duration. This could be explained by the fact that for a given fluence, the lattice temperature is constant whatever the pulse duration, however, the electron temperature is rising to a larger value for shorter pulse duration, according to the three-temperature model [6, 7, 11]. If the two temperatures are very different, more dissipation will take place and consequently more energy is required to reach the threshold equilibrium temperature to allow for a multi-domain state [35].

As shown in Fig. 4(c) and (d), Fig. 6(b) and Fig. S5, $F_{Switch}^{Min}$ does increase if $T_C$ increases but it is unclear whether it is due to the fact that $T_C$ increases with Co concentration or because $T_C(GdCo) > T_C(GdTbCo) > T_C(GdDyCo) > T_C(GdHoCo)$ for a given Co and Gd concentration, as previously mentioned [36]. Unfortunately, we are unable to qualitatively determine the Curie temperature of our samples. Firstly, because of our limited temperature range, but more importantly because the sample properties would be affected irreversibly by applying temperatures above 400K (loss of perpendicular anisotropy). In Fig. 4(b) and 6(b), if we consider that for a given Co concentration, $T_C$ is kept constant, $F_{Switch}^{Min}$ decreases with increasing Gd due to the amount of spin current from the Gd demagnetization ($j = \frac{dM_{Gd}}{dt}$) increasing. Fig. S5 demonstrates that the minimum $F_{Switch}^{Min}$ is obtained for low Tc and high Gd concentration. It also explains why if the pulse duration increases, $F_{Switch}$ increases ($\frac{dM_{Gd}}{dt} \sim \frac{M_{Gd}}{\tau}$). It is amazing that even a Gd concentration as small as 1.5 % is enough to switch magnetization but in this case $F_{Switch}^{Min}$ is large and switching is only observed if τ is small and $F_{Multi}$ is large. It is worth addressing that the lowest Gd concentration of 1.5 % is acquired when the RE-TM alloys are near the compensation point. Highlighting this issue is helpful in understanding the criteria ($F_{Switch}$ (τ) < $F_{Multi}$ (τ)) proposed in this work. For the samples near the compensation point, the value of $F_{Multi}$ is high, meaning we do not need a huge amount of Gd to reduce the value of $F_{Switch}$ for AO-HIS to be observed. On the other hand, once the sample is away from the compensation point, we can clearly see that a higher Gd concentration is necessary to observe AO-HIS (guided by the two arrows in Fig. 1(a)). It is because more Gd leads to lower $F_{Switch}$ and this is certainly required when $F_{Multi}$ is reduced as the sample is away from the compensation.

In Fig. 6(c) and (d), we explicitly show the evolution of both the slope $k$ and $\tau_{max}$ as a function of Co and Gd concentrations. The slope $k$ is also a reflection of how easy it is to switch magnetization as one can deduct from Fig. 4(c): low values of $F_{Switch}^{Min}$ give rise to large $k$ values and vice versa. From the above discussions, our study suggests that $\tau_{max}$ is just a consequence of the crossing between $F_{Switch}^{Min}$ and $F_{Multi}$. Large $\tau_{max}$ is then obtained when both the slope of $F_{Switch}$ (τ) ($k$) and $F_{Multi}$ are large.

## 5. Field-dependent magnetization switching and time-resolved measurements

The above results suggest that AO-HIS could be observed on much larger alloy composition and time ranges if the stray field would not break the magnetic configuration into domains, preventing the observation of AO-HIS. To prove this point, we decided to carry out two experiments. In the first one, a perpendicular magnetic field is applied to compete with the stray field. In the second one, since we expect the effect of the dipolar field to be felt at long time scales, we carried out a time dependent measurement hoping to observe AO-HIS at short time scales and a multi-domain formation at longer time scales.

As shown in Fig. 7, for a 10 nm-$Gd_{18}Dy_8Co_{74}$ sample when the laser fluence F > $F_{Multi}$, the multidomain structure occurs in the middle of the spot. As a result, for no applied magnetic field, three different regions are defined with increasing distance $r$ from the center of the spot, i.e., multi-domain ($0 < r < r_1$), switching ($r_1 < r < r_2$) and no switching ($r > r_2$). When a magnetic field much lower than the coercive field, Hc = 70 mT, varying from + 2 mT to -2mT is applied, for $r > r_1$, no effect of the magnetic field is observed. On the contrary, we found a strong effect of magnetic field on the multi-domain state at the center. The normalized magnetic contrast is around -1 when the applied field is +2 mT opposite to the -80 mT magnetic field applied before the laser pulse. By decreasing the external field to zero, the normalized magnetization tends to approach zero, which corresponds to the random multidomain state. Furthermore, it becomes positive, even saturated (+1), when a high enough negative field is applied. This evolution of the multi-domain state as a function of external field suggests that the dipolar field plays a key role in the formation of the multidomain state.

The above results tend to indicate that such a stray field effect is dominant at very long timescales (quasi-equilibrium state) after excitation by laser pulse. Nonetheless, it is still unclear if the magnetization can be reversed at ultrashort timescales, even in the presence of a stray field strong enough to generate multidomain states at equilibrium. We then studied the time-resolved dynamics for a 10nm-$Gd_{13}Dy_7Co_{80}$ alloy, which shows only multi-domain states after single pulse measurements (Fig. 1(a)). The multi-domain structure obtained 10 second after a single 4.5 mJ/cm$^2$ laser pulse is shown in Fig. 8(a). Interestingly, for the picture showing the magnetic state at 10 ps, a reversal of magnetization indicated by the negative value of the normalized magnetic contrast is observed. In Fig. 8(b), the time-resolved magnetization dynamics for a series of laser fluences demonstrate that the magnetization is switched within less than 10 ps, although a multi-domain state is observed for this sample at long timescales.

We could then conclude that in the above case and for all measured samples the limiting factor for observing AO-HIS is the multi-domain formation occurring at longer time scales. One of the remaining questions and certainly one the most important is then 'Why is Gadolinium so important [37], even when present in such small quantities?'. One could argue that with Gd being an *s* ion with a low spin-orbit coupling, the angular momentum generated by the Gd demagnetization is less likely to be absorbed by the lattice and thus more efficient for generating the switching [38,39]. Choi *et al*. [34] have also demonstrated that in GdFeCo alloys the spin accumulation generated by the Gd sublattice is large and is present 3 ps after the exciting laser pulse, and after the spin accumulation generated by the transition metal sublattice at around 1 ps. The time delay before the Gd relaxation may also be an important feature leaving time for the rest of the sample to demagnetize before absorbing the angular momentum released from the Gd. Finally, considering the "Dieke diagram" presented and explained in Ref [40] for the different rare-earth systems, it seems that for gadolinium there are no excited states available which could be reached with the energy provided by the laser pulse [41]. The later could explain a more efficient angular momentum transfer for Gd.

From the above discussions, several possibilities could be envisioned to increase the variety of material demonstrating AO-HIS and allow the observation for larger pulse durations (i.e., increasing $\tau_{max}$). The first idea is to study samples with in plane magnetization to avoid the generation of multi-domain state-inducing stray fields. In this case, $F_{Multi}$ will not be a limiting factor. $F_{Switch}$ should then only be limited by a "real" demagnetization or the laser damaging the sample. This is what Lin *et al*. tend to demonstrate in their current work [42]. Another way of limiting the effect of $F_{Multi}$ while keeping the PMA materials is to study ultra-thin samples [43]. Finally, another idea is to include a heat sink in the sample so that the heat is dissipated before the dipolar or stray fields are able to act on the magnetization to generate the dipolar field. This is what Verges *et al*. [35] are showing.

### III. Conclusions

In this study, the single-shot AO-HIS state diagrams for various $Gd_yRE_{1-x-y}Co_x$ (RE = Dy, Tb, Ho) alloys are systemically presented. We demonstrated that while in all those alloys, Gd is required, this can be at a concentration as low as 1.5% to still observe AO-HIS. We have determined the criteria needed to be fulfilled to observe AO-HIS by studying the evolutions of two threshold fluences: the switching fluence ($F_{Switch}$) which leads to AO-HIS and the multi-domain fluence ($F_{Multi}$) which leads to a multi-domain state generated by the stray field present in a PMA thin film. $F_{Switch}$ is found to be given by the amount of Gd present and the Curie temperature $T_C$ of the layer, while $F_{Multi}$ is determined by the saturated magnetization, $M_s$, the exchange, $J_{eff}$, and the anisotropy constant, $K_u$. Considering that $F_{Switch} < F_{Multi}$ is a necessary condition to observe AO-HIS, we could provide practical ways to enlarge the region of AO-HIS, including both the materials compositions and the laser parameters. From those conclusions, we could propose several types of heterostructures demonstrating AO-HIS which will be of great significance for the future of ultra-fast spintronic applications.

**Acknowledgements**

The authors thank Eric Fullerton, Mike Coey and Bert Koopmaans for fruitful discussions. This work is supported by the ANR-20-CE09-0013 UFO, the Institute Carnot ICEEL for the project "CAPMAT" and FASTNESS, the Région Grand Est, the Metropole Grand Nancy, for the Chaire PLUS by the impact project LUE-N4S, part of the French PIA project "Lorraine Université d'Excellence" reference ANR-15-IDEX-04-LUE, the "FEDERFSE Lorraine et Massif Vosges 2014-2020" for IOMA a European Union Program, the European Union's Horizon 2020 research and innovation program COMRAD under the Marie Skłodowska-Curie grant agreement No 861300, the ANR project ANR-20-CE24-0003 SPOTZ. This article is based upon work from COST Action CA17123 MAGNETOFON, supported by COST (European Cooperation in Science and Technology). All fundings were shared equally among all authors. W. Z. gratefully acknowledges the National Natural Science Foundation of China (grant no. 12104030), the China Postdoctoral Science Foundation (grant no. 2022M710320) and the China Scholarship Council.

**APPENDIX: METHODS AND MATERIALS**

**1. Film Growth**

We prepared a large number of 10 nm-thick $Gd_yRE_{1-x-y}Co_x$ (RE = Dy, Tb, Ho) thin films by magnetron sputtering. Thin film hetero-structures are prepared onto a Corning 1737 float glass substrate according to the following multilayered structure: glass/Ta(5 nm)/Pt(3 nm)/$Gd_yRE_{1-x-y}Co_x$ (10 nm)/Pt(5 nm). The base pressure was $< 2\times10^{-8}$ mbar, and a working pressure of Ar at $2\times10^{-3}$ mbar during sputtering was maintained.

**2. AO-HIS measurements**

A Ti: sapphire fs-laser source with a regenerative amplifier was used to generate the linearly polarized pump laser beam with a central wavelength of 800 nm (1.55 eV) and a 5 kHz repetition rate. To record the sample's magnetic state after a single ultra-short laser pulse, a Kerr microscope is used with an LED with a central wavelength around 630 nm as the light source. Pictures are taken before and after each laser shot, to allow a background image correction. To ensure a well-defined initial state, the film is saturated using an external magnetic field whose exact value depends on the coercivity of the samples.

**3. Pump-probe magnetization dynamics measurements**

Time-resolved experiments on continuous 10nm-$Gd_{13}Dy_7Co_{80}$ alloy were conducted with a Yt fiber laser PHAROS from Light Conversion with a regenerative amplifier with a repetition rate fixed at 5 kHz. Two optical parametric amplifiers (OPA) were used to tune the wavelengths of the pump and probe beams at 800 nm and 650 nm with pulse durations of 100 fs and 178 fs, respectively. A polarizer was coupled to a half-wave plate to be able to control the pump power. The spot size of the writing pulse was about 225 μm. Images have been recorded for any pump-probe delay by a CCD (EO-5023M MONO USB CCD camera, exposure time 285.9 ms, 3.65 frames/s, accumulation of 10 images) as tiff files. A reset magnetic field around 100 Oe is applied during the time-resolved measurements.

# Bibliography

[25] C. Kooy and U. Enz, *Experimental and theoretical study of the domain configuration in thin layers of BaFeO,* Philips Res. Rep. 15, 7 (1960).

[26] O. Hellwig, A. Berger, J. B. Kortright and E. E. Fullerton, *Domain structure and magnetization reversal of antiferromagnetically coupled perpendicular anisotropy films,* J. Magn. Magn. Mater. 319 13–55 (2007).

[27] M. Salah El Hadri, M. Hehn, P. Pirro, C-H. Lambert, G. Malinowski, E. E. Fullerton, and S. Mangin, *Domain size criterion for the observation of all-optical helicity-dependent switching in magnetic thin films*, Phys. Rev. B. 94, 064419 (2016).

[28] A. Westover, K. Chesnel, K. Hatch, P. Salter, O. Hellwig, *Enhancement of magnetic domain topologies in Co/Pt thin films by fine tuning the magnetic field path throughout the hysteresis loop*, J. Magn. Magn. Mater 399, 164–169 (2016).

[29] M. Beens, KA. de Mare, RA. Duine, B. Koopmans, *Spin-polarized hot electron transport versus spin pumping mediated by local heating,* Journal of Physics: Condensed Matter 35 (3), 035803 (2022).

[30] Q. Remy, J. Igarashi, S. Iihama, G. Malinowski, M. Hehn, J. Gorchon, J. Hohlfeld, S. Fukami, H. Ohno, S. Mangin, *Energy Efficient Control of Ultrafast Spin Current to Induce Single Femtosecond Pulse Switching of a Ferromagnet*, Adv. Sci. 7, 2001996 (2020).

[31] S. Iihama, Y. Xu, M. Deb, G. Malinowski, M. Hehn, J. Gorchon, E. E. Fullerton, and S. Mangin, *Single-Shot Multi-Level All-Optical Magnetization Switching Mediated by Spin Transport,* Adv. Mater. 30, 1804004 (2018).

[32] J. Igarashi, W. Zhang, Q. Remy1, E. Díaz, J. X. Lin, J. Hohlfeld, M. Hehn, S. Mangin, J. Gorchon, G. Malinowski, *Optically induced ultrafast magnetization switching in ferromagnetic spin valves*, Nat. Mater. 22, 725 (2023).

[33] Q. Remy, J. Hohlfeld, M. Vergès, Y. Le Guen, J. Gorchon, G. Malinowski, S. Mangin and M. Hehn, *Accelerating ultrafast magnetization reversal by non-local spin transfer,* Nat. Commun. 14, 445 (2023).

[34] G. M. Choi and B. C. Min, *Laser-driven spin generation in the conduction bands of ferrimagnetic metals,* Phys. Rev. B. 97, 014410 (2018).

[35] M. Verges et al. In preparation. (2023)

[36] J. Stohr and H. C. Siegmann, Magnetism: From fundamentals to nanoscale dynamics. (Higher education press, 2012).

**Figure captions:**

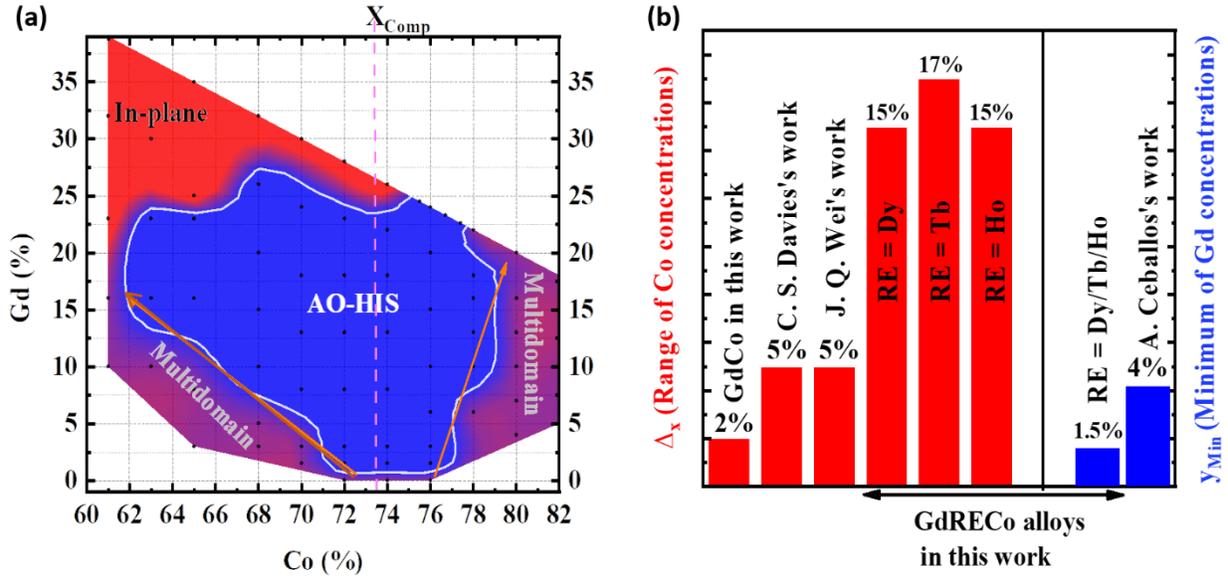

FIG. 1. (a) Mapping of the magnetic states after one single ultra-short laser pulse with 50 fs duration as a function of the Gd and the Co concentration for 10 nm thick $Gd_yRE_{1-x-y}Co_x$ samples with RE = Dy. Depending on the concentrations, three regions can be defined as *In-plane*, *AO-HIS*, and *multidomain states*. The vertical pink dash line represents the compensation concentration. The two solid arrows are a guide to the eyes between AO-HIS and multidomain states. (b) Summary of the Co composition range ($\Delta x$) and minimum Gd compositions needed to observe single single-short laser pulse AO-HIS for various studies (C.S. Davies [9], J.Q. Wei [16] A. Ceballos [23]).

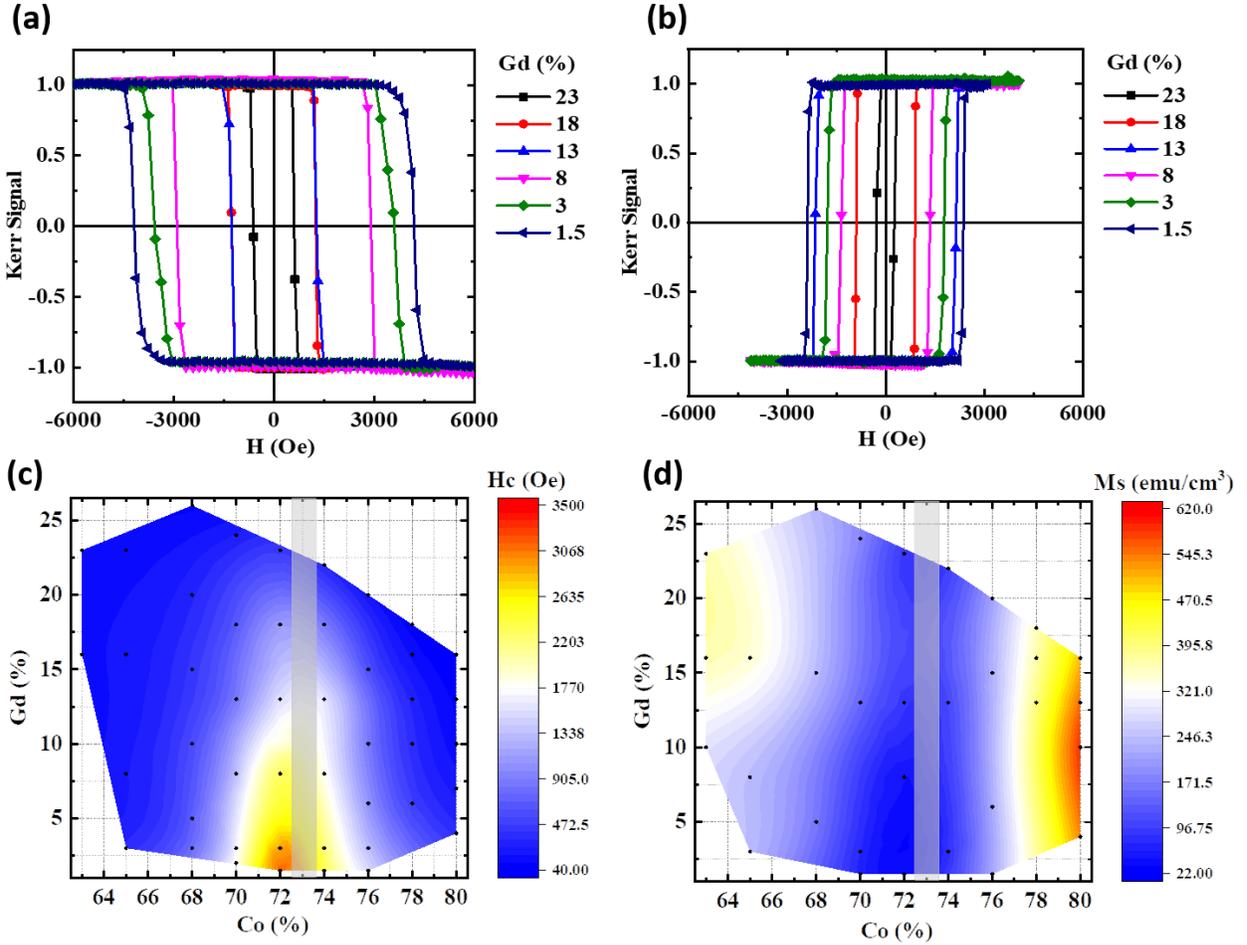

FIG. 2. Static magnetic properties of 10 nm $Gd_yDy_{1-x-y}Co_x$ films. (a) and (b) Normalized Magneto-Optical Kerr Effect (MOKE) as a function of the magnetic field H applied perpendicular to the film plane, with x = 72%, and 74%, respectively. The hysteresis loops are inverted for a and b demonstrating that the magnetization compensation is obtained for a Co concentration between 72%, and 74%, for all Gd concentration. (c) and (d) Mapping of the coercivity and the saturated magnetization as a function of Gd and Co concentration. Dots indicate measured data.

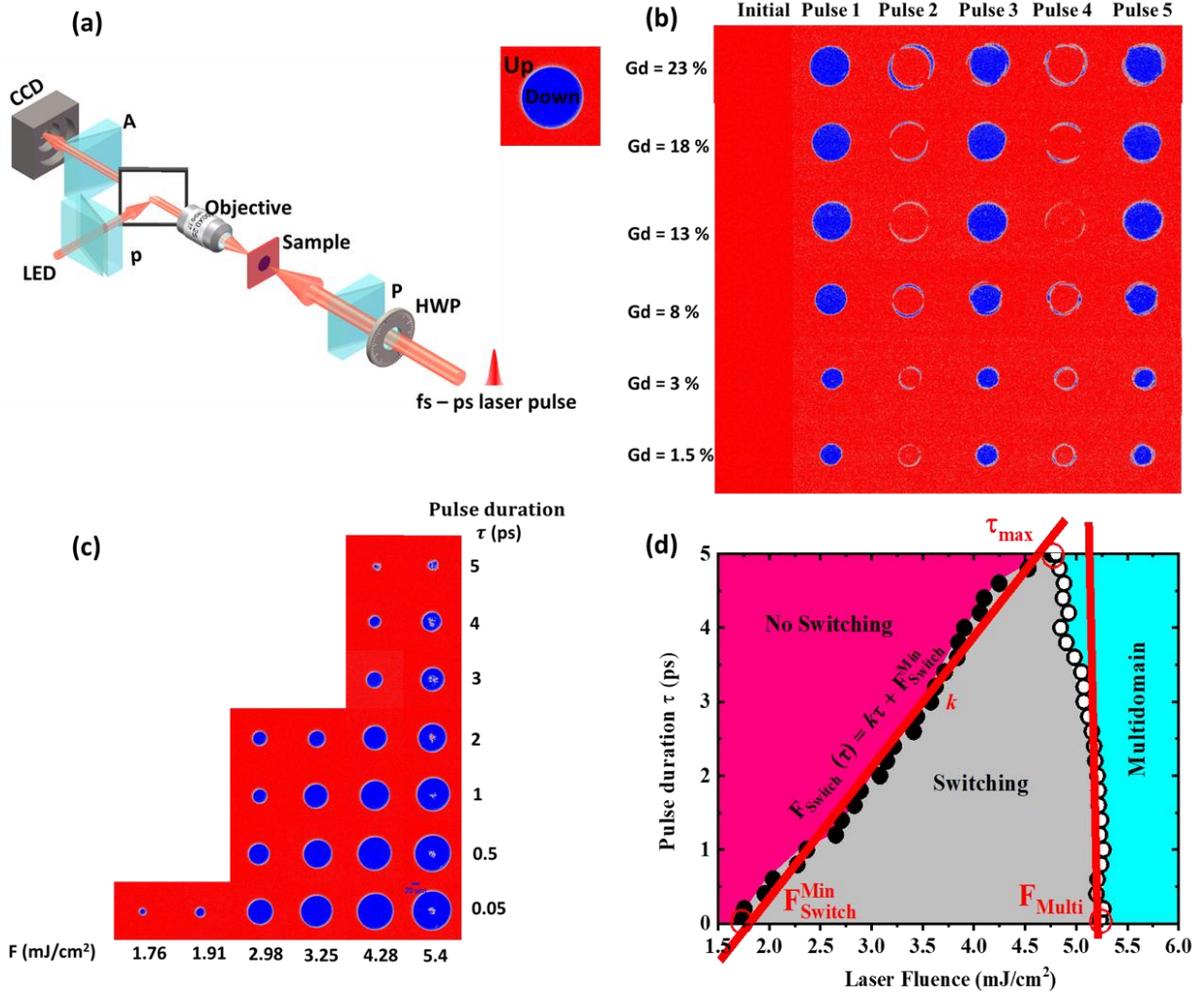

FIG. 3. Single pulse all optical magnetization toggle switching and construction of AO-HIS state diagrams. (a) Sketch of the MOKE microscopy setup using single femtosecond laser pulse excitations. HWP: Half wave-plate; P: Polarizer; A: Analyzer; LED: Light emitting diode; CCD: Charge coupled device. (b) Magneto-optical images of 10 nm $Gd_yDy_{1-x-y}Co_x$ alloys with x = 72% and y varying from 1.5% to 23% after exposure to a sequence of single linearly-polarized laser pulses with a pulse duration of τ = 50 fs and a fluence F=3.2 mJ/cm$^2$. (c) Magneto-optical images of 10 nm $Gd_yDy_{1-x-y}Co_x$ alloys with x = 72%, y = 23%, after exposure to a single linearly-polarized laser pulse with various pulse durations (τ) and laser fluences (F). (d) Details of the state diagram constructed for $Gd_{23}Dy_5Co_{72}$: switching fluence $F_{Switch}$ (solid black dot) and multidomain fluence $F_{Multi}$ (open black dot) as a function of the pulse durations (τ). Three regions are defined by the evolution of $F_{Switch}$ (τ) and $F_{Multi}$ (τ): switching, no switching, and multidomain states. $F_{Switch}^{Min}$, $k$ and $\tau_{max}$ are defined by assuming $F_{Switch}(\tau)= k\tau + F_{Switch}^{Min}$ and $F_{Switch}(\tau_{max})= F_{Multi}(\tau_{max})$.

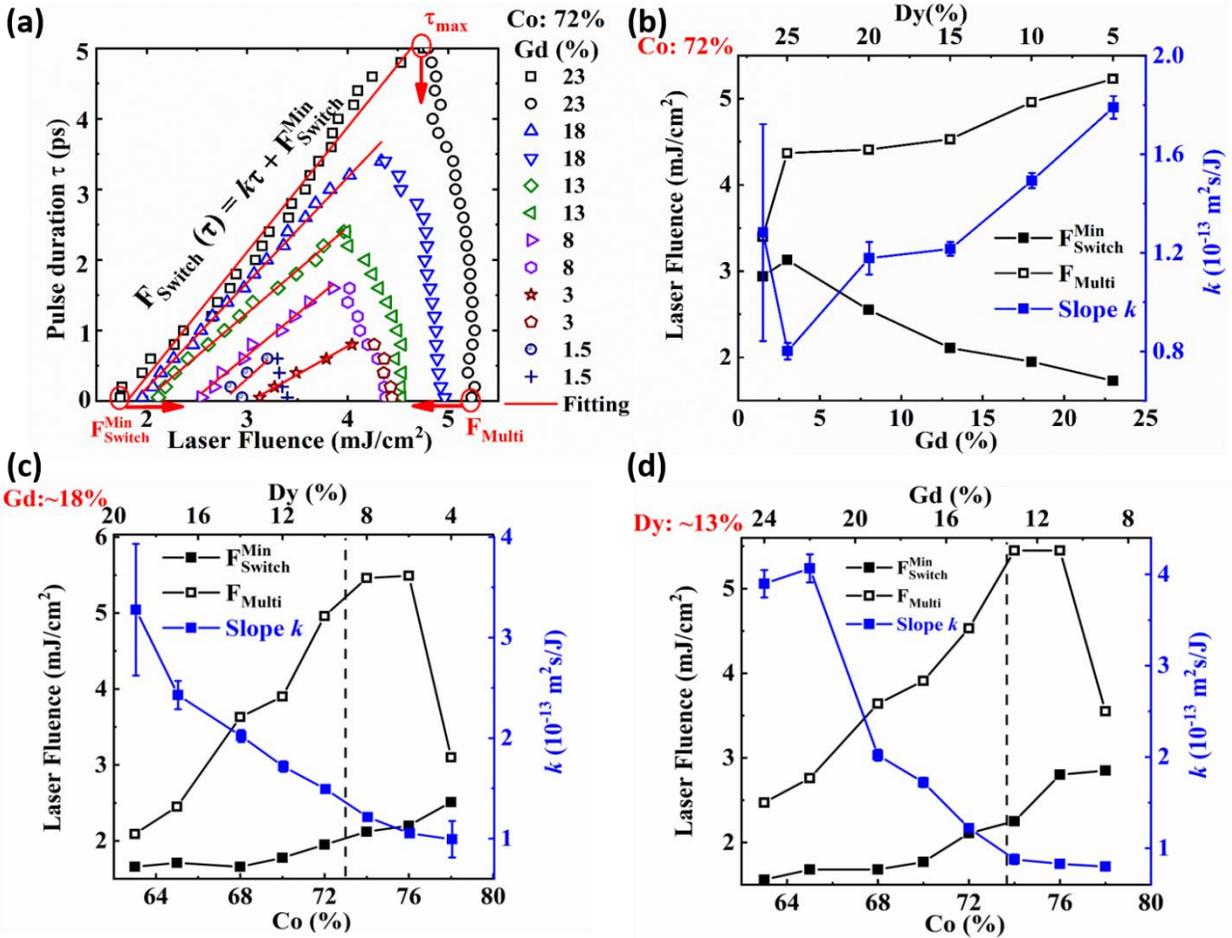

FIG. 4. Evolution of AO-HIS state diagrams for a 10 nm $Gd_yDy_{1-x-y}Co_x$ alloy sample. (a) Evolution of the magnetic state diagrams for various Gd concentrations while the Co concentration is kept constant at x= 72%, $Gd_yDy_{0.28-y}Co_{0.72}$. The parameters $F_{Switch}$, $F_{Multi}$, and $k$ are extracted from Fig. 4(a) and Fig. S4(a) and (b), as a function of Gd and Co concentrations (b) for a constant Co concentration equal to 72% (c) for a constant Gd concentration equal to 18% (d) for a constant Dy concentration equal to 13%. The dashed lines in (c) and (d) correspond to the compensation point.

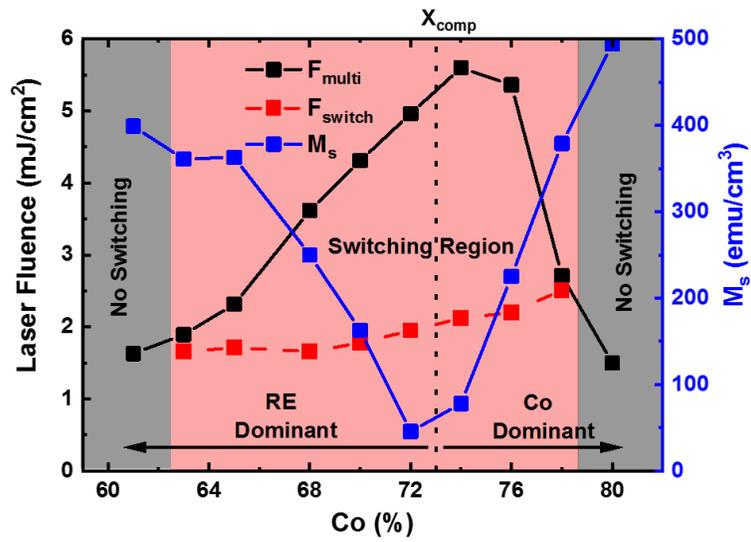

FIG. 5. Evolution of $F_{Multi}$, $F_{Switch}$, and Ms as a function of the Co compositions for a 10 nm GdDyCo alloys, with a constant Gd concentration equal to 18%. $F_{Multi}$ and $F_{Switch}$, are extracted from the AO-HIS experiments using a 50fs pulse duration.

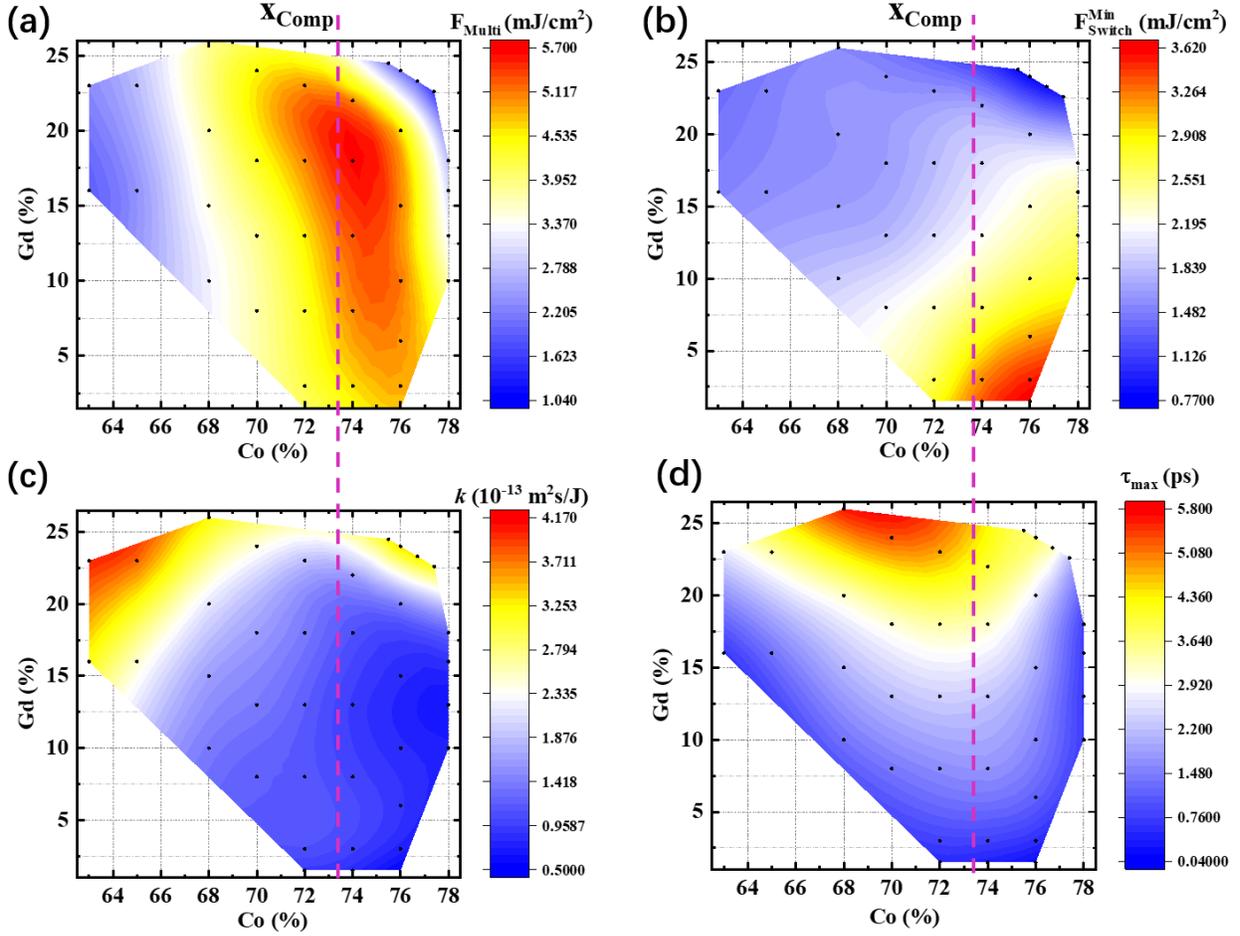

FIG. 6. Mapping of laser parameters allowing for optical magnetization switching as a function of the Gd and Co concentration for a 10 nm GdDyCo alloys. (a) The threshold laser fluence to observe multi-domain state $F_{Multi}$ (b) The threshold laser fluence to observe AO-HIS $F_{Switch}$, (c) the maximum pulse duration to observe AO-HI $\tau_{max}$ and (d) $k$, the slope describing the dependence of $F_{Switch}$ on pulse duration. The dashed lines correspond to the compensation composition of the alloys. Dots indicate measured data.

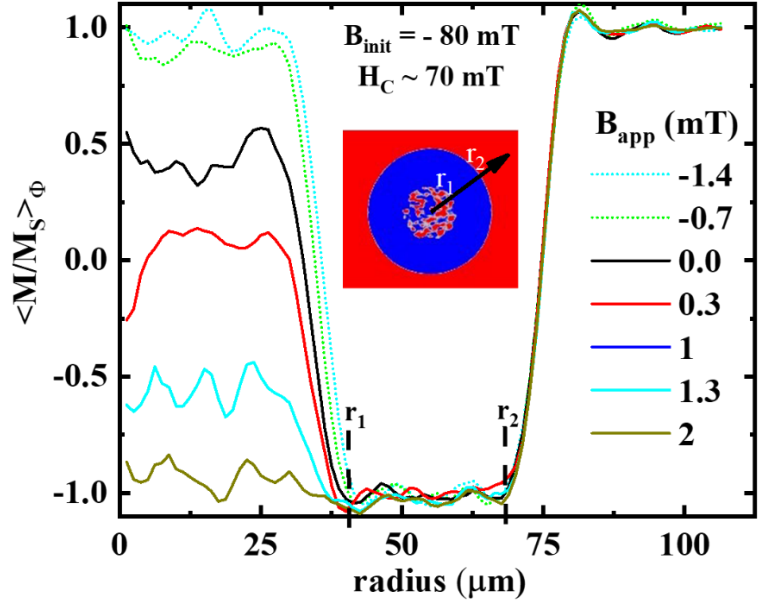

FIG. 7. Influence of applied magnetic fields on the multidomain state after a single laser pulse for a 10nm-$Gd_{18}Dy_8Co_{74}$ sample. Before the single pulse measurements, the sample is saturated under a -80 mT field. During the laser irradiation, various external fields (-1.4, -0.7, 0, 0.3, 1, 1.3 or 2 mT) were applied. For zero applied field after a single laser pulse: multidomain structures appear for $r < r_1$, magnetization reversal for $r_1 < r < r_2$, and no magnetization change for $r > r_2$.

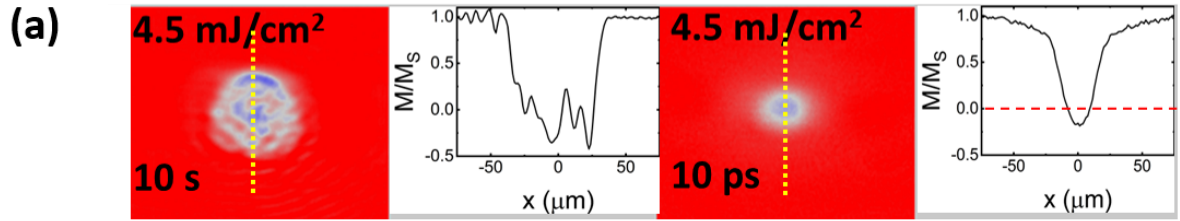

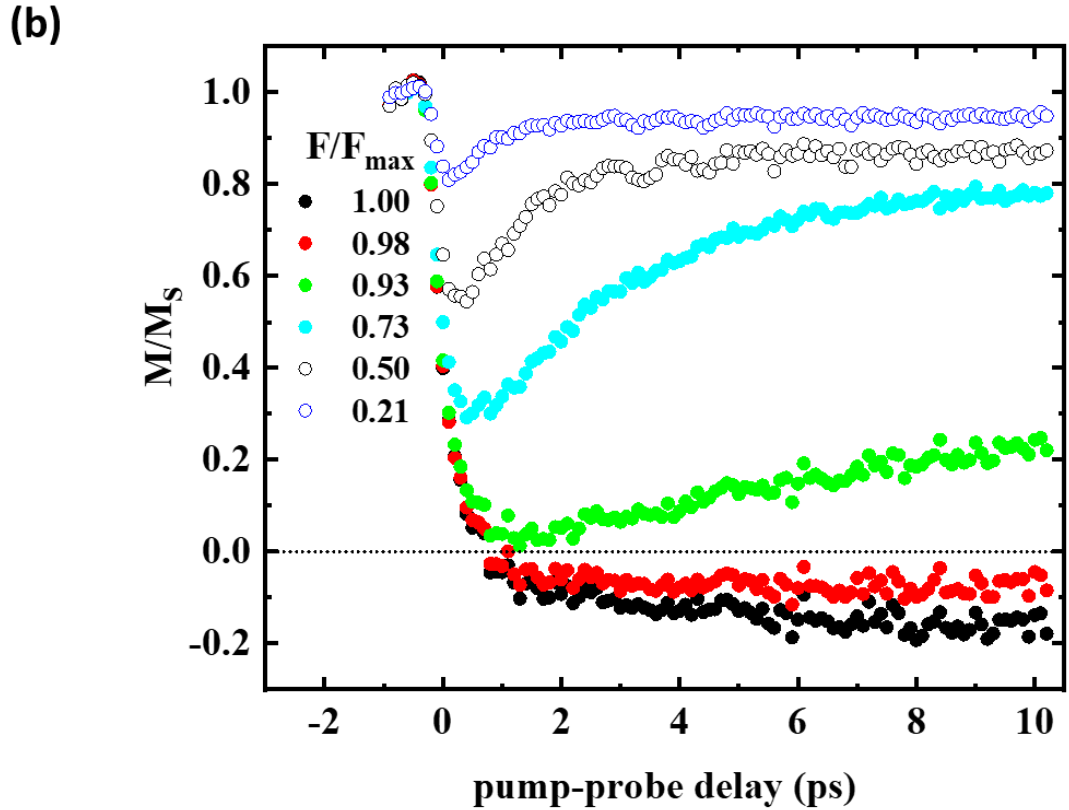

FIG. 8. Time-resolved magnetization switching dynamics on of 10nm-$Gd_{13}Dy_7Co_{80}$ alloy sample. (a) images obtained 10 seconds after shining the laser pulse and 10 pico-seconds after shining the laser pulse. (b) Evolution of the normalized magnetization as a function of time. The sample's magnetization is reversed within less than 10 ps for proper laser fluences.